\documentclass[aps,pra,twocolumn,groupedaddress,showpacs]{revtex4}
\usepackage{amsmath,amsfonts,amssymb,bbm,citesort,graphicx}

\newcommand{\eqn}[1]{\begin{equation} #1 \end{equation}} 
\newcommand{\aln}[1]{\begin{align} #1 \end{align}}       
\newcommand{\mc}{\mathcal}                               
\newcommand{\mbf}{\mathbf}                               
\newcommand{\eq}[1]{(\ref{#1})}              
\newcommand{\wtilde}{\widetilde}             
\newcommand{\wbar}{\overline}                
\renewcommand{\l}{\left}                     
\renewcommand{\r}{\right}                    

\begin{document}


\title{Spacing statistics in two-mode random lasing}


\author{Oleg Zaitsev}
\email[E-mail: ]{oleg.zaitsev@uni-duisburg-essen.de}
\affiliation{Fachbereich Physik, Universit\"at Duisburg-Essen, 
             Lotharstr.~1, 47048 Duisburg, Germany}


\date{October 19, 2007}

\begin{abstract}

The distribution of spacings between the lasing frequencies for an ensemble of
random lasers in the two-mode regime was computed. The random lasers are
implemented as open chaotic cavities filled with an active medium. The spectral
properties of the passive cavities are modeled with non-Hermitian random
matrices. The spacing distribution is found to depend on the relation between
the gain-profile width and the mean spacing of the passive-cavity modes. The
distribution displays mode repulsion and, under certain conditions, agrees with
the Wigner surmise. The role of mode competition is discussed. 

\end{abstract}


\maketitle


\section{Introduction}

The term ``random laser'' usually refers to lasing systems based on disordered
materials or substantially open wave-chaotic resonators~\cite{cao03}. They are
distinguished from conventional lasers by leaky modes (because of the absence
or openness of the resonator) and almost random field distributions (due to
disorder or chaotic shape). Coherent lasing in disordered materials has been
observed via emission spectra~\cite{cao99} and photon-count
measurements~\cite{cao01}. The role of strong localization of light in
disordered lasers remains an open theoretical question~\cite{vann01,apal02}. 
In the recent work on the chaotic-laser theory~\cite{hack01,hack03}, the
standard laser models~\cite{sarg74,hake85} were extended to cavities with
spectrally overlapping modes and equipped with the ideas of quantum
chaos~\cite{haak01}. The spatial structure of lasing modes was also
considered~\cite{ture06}. 

A substantial aspect of the probabilistic description of random lasers is the
mode statistics. For example, the mean number of lasing modes in
weakly~\cite{misi98} and strongly~\cite{hack05} open resonators, as well as
the variance~\cite{zait06}, were calculated with the help of random-matrix
theory~\cite{fyod97}. The present study was partly motivated by recent
experiments in a porous-GaP laser~\cite{mole06}. The observed distribution of
spectral mode spacings could be well fitted with the Wigner surmise for the
Gaussian orthogonal ensemble (GOE) of random matrices~\cite{haak01}. As a
first step toward understanding of the multimode regime, here I propose a
theory of spacing statistics for two-mode lasing.  The calculation is based on
the random-matrix approach of Refs.~\cite{hack05,zait06}. The results point to
the spectral repulsion of lasing modes on the scale inherited from the passive
cavity. A different repulsion mechanism, requiring inhomogeneous broadening,
was suggested in Ref.~\cite{cao03b}.

\section{Theory}

\subsection{Laser equations}

As a model of a random laser we adopt an open resonator of irregular shape,
having almost random eigenfunctions~\cite{berr77}. The lasing takes place due
to a pumped active medium inside the resonator. Both the field and the active
atoms interact with the environment. The environment variables can be
eliminated from the equations of motion by inclusion of the effective damping
and noise terms~\cite{sarg74}. The number of photons in a mode~$k$ is
controlled via the bosonic operators $a_k$ and~$a_k^\dag$. The occupation of
the two active levels of an atom~$p$ will be described by the pseudospin-$1/2$
operators $\sigma_p$ and $s_p = s_p^\dag$ satisfying the commutation relation
$\l[\sigma_p, s_p \r] = \sigma_p$. The operator~$\sigma_p$ transfers the
electron from the upper to the lower level and $s_p$ yields the population
difference between the levels. We will work in the classical approximation,
replacing operators with $c$~numbers and neglecting the noise. Then $I_k \equiv
|a_k|^2$ will be proportional to the mode intensity and $\sigma_p$ will
characterize the atom's polarization. The coupled classical equations of motion
(equivalent to the Maxwell-Bloch equations) take the form
\aln{
  &\dot a_k = - \l(i \omega_k + \kappa_k \r) a_k + g \sum_{p'} \wtilde
  \phi_k^* \l(\mbf r_{p'} \r) \sigma_{p'}, 
  \label{dota}\\
  &\dot \sigma_p = - \l(i \nu + \gamma_\perp \r) \sigma_p + 2g \sum_{k'}
  \phi_{k'} \l(\mbf r_p \r) a_{k'} s_p,
  \label{dotsig}\\
  &\dot s_p = \gamma_\parallel \l(S - s_p \r) - g \sum_{k'} \l[\phi_{k'}
  \l(\mbf r_p \r) a_{k'} \sigma_p^* + \text{c.c.} \r].
  \label{dots}
}
Here $\omega_k - i \kappa_k$ are the complex frequencies of passive modes in
the cavity, $\nu$~is the atomic transition frequency, $\gamma_\perp$~and
$\gamma_\parallel$ are the polarization and inversion decay rates, and
$S$~specifies the pump strength. The atom-field coupling is measured by the
parameter $g \simeq d \sqrt{2 \pi \nu / \hbar}$, where $d$ is the dipole
moment for the atomic transition. All modes~$k$ are assumed to be linearly
polarized in the same direction. The field amplitude, evaluated at the atom
position $\mbf r_p$, is described by the normalized wavefunction~$\phi_k (\mbf
r)$. As an eigenfunction of a non-Hermitian operator, it has an adjoint
eigenfunction~$\wtilde\phi_k (\mbf r)$ (Cf.\ Ref.~\cite{sieg89a}). The
functions $\phi_k$ and $\wtilde\phi_{k'}$ are orthogonal for $k \ne k'$
(biorthogonality). Equations \eq{dota}-\eq{dots} are written in the
rotating-wave approximation. It neglects antiresonant products of the type
$a_k \sigma_p$, oscillating with a double optical frequency.

\subsection{Lasing frequencies}

To solve Eqs.\ \eq{dota}-\eq{dots}, we assume that each lasing mode $a_k (t) =
\alpha_k (t) \exp (-i \Omega_k t)$ oscillates with a constant frequency
$\Omega_k$ and a slowly varying amplitude~$\alpha_k (t)$. Following a standard
procedure~\cite{hake85} we expand $\sigma_p (t) = \sum_k \sigma_{pk} (t) \exp
(-i \Omega_k t)$ and $s_p (t) - S = \sum_{kk'} \l\{s_{pkk'} (t) \exp \l[-i
\l(\Omega_k - \Omega_{k'}\r) t\r] + \text{c.c.} \r\}$ and neglect all other
possible oscillating terms. Furthermore, only contributions up to the order of
$\alpha_k^3$ in $\sigma_p$ and $\alpha_k^2$ in $s_p$ are kept. Under the
assumption $\l|\dot \alpha_k / \alpha_k \r| \ll \gamma_\perp,
\gamma_\parallel$, the atomic variables can be eliminated from the equations
of motion.  The resulting equations for~$\alpha_k$,
\eqn{
  \dot \alpha_k = \alpha_k \l( p_k - q_k I_k - \sum_{k' \ne k} r_{kk'}
  I_{k'}\r),
  \label{req}
}
contain linear and nonlinear (intensity-dependent) terms. The coefficients~are
\aln{
  &p_k = i \l( \Omega_k - \omega_k \r) - \kappa_k + G \mc D_k, \\
  &q_k = 2c B_{kkkk} \mc D_k \mc L_k, \\
  &r_{kk'} = c \mc D_k \l[ 2 B_{kkk'k'} \mc L_{k'} + B_{kk'kk'}
  \mc D_{kk'}^\parallel \l( \mc D_{k'}^* + \mc D_k\r) \r],
}
where $ B_{klmn} \simeq V \int d \mbf r\, \wtilde \phi_k^* (\mbf r)\, \phi_l
(\mbf r)\, \phi_m (\mbf r)\, \phi_n^* (\mbf r)$,  $\mc D_k = \l[1 - i
\l(\Omega_k - \nu \r)/ \gamma_\perp \r]^{-1}$, $\mc D_{kk'}^\parallel = \l[1 -
i \l(\Omega_k - \Omega_{k'} \r)/ \gamma_\parallel \r]^{-1}$, $\mc L_k =
\text{Re}\, \mc D_k$, $G = 2 g^2 S \mc N/ \gamma_\perp V$, $c = 4 g^4 S \mc N/
\gamma_\perp^2 \gamma_\parallel V^2$, $\mc N$~is the number of atoms, and $V$
is the cavity volume. The linear absorption $\kappa_k$ is counteracted by the
linear gain $G \mc L_k$, which has a Lorentzian profile of
halfwidth~$\gamma_\perp$. The chaotic wavefunctions for different modes behave
like independent Gaussian random functions, on scales larger than the
wavelength~\cite{berr77}. This allows one to approximate the correlation
parameters $B_{kkk'k'} \simeq B_{kk'kk'} \simeq 1 + 2
\delta_{kk'}$~\cite{misi98,hack05}. 

In a steady state $\dot \alpha_k = 0$, the bracketed portion of Eq.~\eq{req}
must vanish for all $k$, such that $\alpha_k \ne 0$. This makes a system of
complex algebraic equations, from which the $\Omega_k$'s and $I_k$'s of the
lasing modes can be determined. To use this procedure, however, it must be
known \emph{a priori} which modes have nonzero intensity. If this is not the
case, one has to \emph{assume} that certain modes are lasing, solve the system
of equations, and then verify that all found intensities are positive and the
solution is stable. In view of the complexity of the problem, I restricted the
present study to the case of two-mode lasing. 

If only one mode is lasing, its frequency is~\cite{hake85}
\eqn{
  \Omega_k^{(1)} = \frac {\omega_k + \nu \kappa_k/ \gamma_\perp} {1 + \kappa_k
  / \gamma_\perp}.
  \label{O1}
}
The positive-intensity condition requires that the linear gain for this mode
exceeds the linear absorption. Hence, the actual first lasing mode, labelled
$k=1$, without loss of generality, is the one with the lowest threshold 
\eqn{
  G_1 = \frac {\kappa_1}  {\mc L_1^{(1)}} = \min_k \l(\frac {\kappa_k} {\mc
  L_k^{(1)}} \r), \quad \mc L_k^{(1)} \equiv \bigl. \mc L_k\bigr|_{\Omega_k =
  \Omega_k^{(1)}}.
}
In other words, the first lasing mode $k=1$ emerges at the pump level $G =
G_1$. Unlike~$\Omega_1^{(1)}$, the frequency of the second lasing mode depends
on the mode intensities. In the results for the spacing statistics below, this
frequency is always taken at the threshold. If $k \ne 1$ is the second lasing
mode, Eq.~\eq{req} provides a pair of complex equations, with an unknown $I_1$
and $I_k \to 0^+$. They yield a single nonlinear equation for the frequency
$\Omega_k^{(2)}$ at the threshold,
\aln{
  &\Omega_k^{(2)} - \Omega_k^{(1)} = R_k \l. \frac { \text{Im} \l[\mc
  D_{k1}^\parallel \l(\mc D_1^* + \mc D_k\r) \r] } {1 + \frac{\kappa_k}
  {\gamma_\perp}} \r|_{\Omega_k = \Omega_k^{(2)}}, 
  \label{frshift}\\
  &R_k \equiv \l. \frac {\kappa_k - \l(\Omega_k - \omega_k \r) \frac {\Omega_k
  - \nu} {\gamma_\perp} - G_1} {4 \mc L_1^{(1)} - \text{Re} \l[\mc
  D_{k1}^\parallel \l(\mc D_1^* + \mc D_k\r) \r]} \r|_{\Omega_k =
  \Omega_k^{(2)}}.
  \label{cI1}
}
The stability of the two-mode solution can be checked as described in
Ref.~\cite{sarg74}. Of all stable candidates, the genuine second lasing mode,
labeled $k=2$, minimizes the threshold, i.e.,
\eqn{
  G_2 = \min_{\text{stable } k \ne 1} \l( G_1 + 6 \mc L_1^{(1)} R_k \r).
  \label{G2}
}
The first-mode intensity at the second-mode threshold is $I_1 = R_2/c$. In
order to estimate the importance of the nonlinear effects (mode competition),
we can also define the would-be second mode neglecting the $I_1$ dependence in
Eq.~\eq{req}. This mode, labeled $k=2'$, has the threshold, simply, $G_{2'} =
\min_{k \ne 1}\l( \kappa_k /\mc L_k^{(1)} \r)$ and the
frequency~$\Omega_{2'}^{(1)}$.

\subsection{Random-matrix model}

The working equations \eq{O1}-\eq{G2} use the eigenfrequencies $\omega_k - i
\kappa_k$ of the passive cavity as an input. In order to describe statistical
characteristics of an ensemble of (nonidentical) chaotic lasers, the passive
spectra can be modeled with the eigenvalues of non-Hermitian random matrices. 
Henceforth we will formally set $\nu = 0$, i.e., the frequencies will be
measured with respect to the atomic frequency. Following Refs.\
\cite{fyod97,hack05,zait06}, we associate with each cavity an $L \times L$
matrix $\hat \omega - i \hat \gamma$. Here, the real symmetric matrix $\hat
\omega$ is chosen from the GOE and $\hat \gamma$ is a fixed diagonal matrix
with $M \ll L$ positive and $L - M$ zero eigenvalues. The eigenvalue density of
$\hat \omega$ obeys the Wigner semicircle law $\rho \l(\omega \r) = \pi^{-1}
\sqrt{1 - \omega^2/4}$ (in dimensionless units), which remains approximately
valid for the real parts of the eigenvalues of $\hat \omega - i \hat \gamma$. 
The integer $M$ is interpreted as the number of spectral bands of the outside
field, to which the cavity field is coupled~\cite{hack03}. The bands, called
the coupling channels, are a consequence of the wave quantization due to the
finite size of the cavity opening. They are similar to conductance channels of
a quantum dot connected to ballistic leads. We will use the model of equivalent
channels, when all nonzero eigenvalues of $\hat \gamma$ are equal to some
$\gamma > 0$. According to random-matrix theory, the coupling strength is
characterized by the parameter $2 \pi \rho \l(\omega \r) / \l( \gamma +
\gamma^{-1} \r)$~\cite{fyod97}. Hence, $\gamma$~between $0$ and~$1$ spans the
whole coupling range from weak to strong. In order to exclude the effects of
variable density~$\rho (\omega)$, only the eigenvalues near the top of the
Wigner semicircle were taken into account in the numerical simulations. That
is, for each random matrix $\hat \omega - i \hat \gamma$, I~used $L_0 \approx
0.36 L$ eigenvalues with the smallest $\l|\omega_k \r|$, thereby allowing for a
4\%~variation of~$\rho \l(\omega_k \r)$.

\section{Numerical results and discussion}

\begin{figure}[tb]
  \centering{\includegraphics[width= .86\linewidth, angle=0]%
  {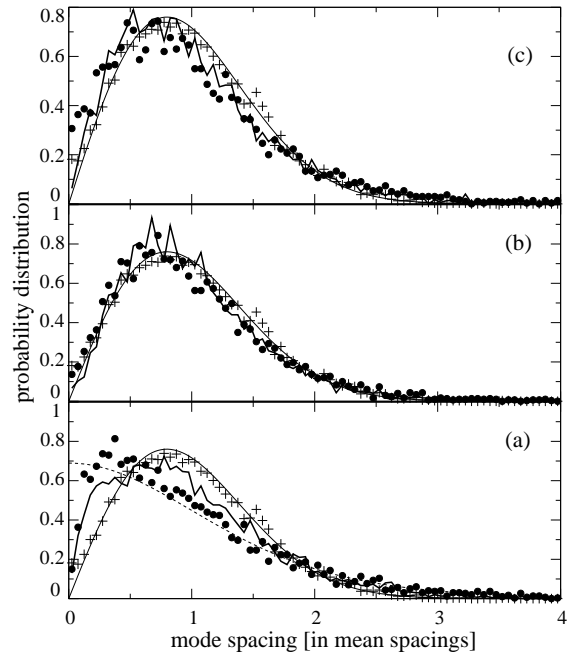}} 
  \caption{\small Distributions of spacings between the first two lasing modes
  for an ensemble of 6,000 non-Hermitian random matrices of size $L = 200$ with
  $M = 6$ coupling channels and coupling parameter $\gamma = 0.1$. In each
  matrix, $L_0 = 72$ eigenvalues closest to the top of the Wigner semicircle
  were taken into account. The average damping $\wbar{\kappa_k} = 3.0 \times
  10^{-3}$ was determined numerically. Shown are the distributions $P \l(\Delta
  \Omega\r)$ (bold solid line), $P' \l(\Delta \Omega'\r)$ (circles) with the
  mode competition neglected, $p \l(\Delta \omega\r)$ (pluses) for the passive
  frequencies, the Wigner surmise~$p_W \l(\Delta \omega\r)$ (thin solid line),
  and the approximation~\eq{Pappr} for~$P' \l(\Delta \Omega'\r)$ (dashed line).
  The spacings are computed in units of the mean values for each distribution
  (Table~\ref{sp}). The gain-profile halfwidth $\gamma_\perp$ is $10^{-1}$~(a),
  $10^{-2}$~(b), and $10^{-3}$~(c). The inversion decay rate~$\gamma_\parallel
  = \gamma_\perp$.}
  \label{p1}
\end{figure}

\begin{figure}[tb]
  \centering{\includegraphics[width= .86\linewidth, angle=0]%
  {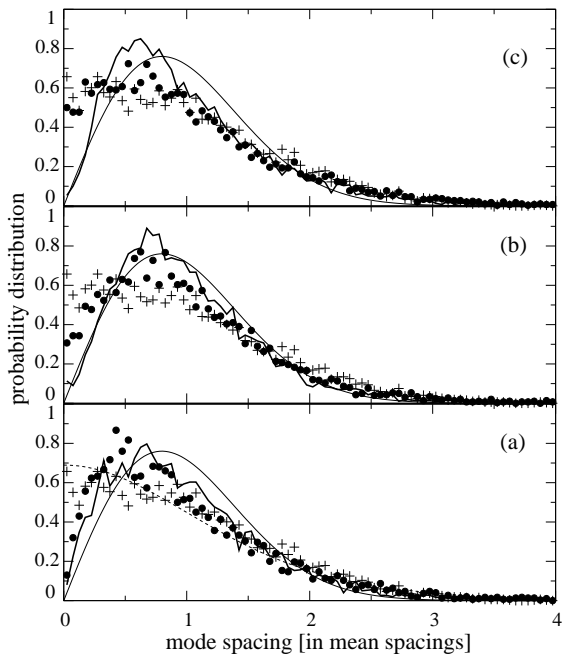}} 
  \caption{\small Same as in Fig.~\ref{p1}, but with $\gamma = 1.0$,
  $\wbar{\kappa_k} = 4.2 \times 10^{-2}$.}
  \label{1p}
\end{figure}

Numerical results for the mode-spacing statistics are shown in Figs.~\ref{p1}
and~\ref{1p}. The first two lasing frequencies $\Omega_1^{(1)}$
and~$\Omega_2^{(2)}$, as well as the second frequency~$\Omega_{2'}^{(1)}$
without mode competition, were computed for each random matrix in an ensemble.
The plots show the spacing distributions $P \l(\Delta \Omega\r)$, $P'
\l(\Delta \Omega'\r)$ (the primed function is not a derivative), and~$p
\l(\Delta \omega\r)$, where $\Delta \Omega \equiv \l|\Omega_1^{(1)} -
\Omega_2^{(2)} \r|$, $\Delta \Omega' \equiv \l|\Omega_1^{(1)} -
\Omega_{2'}^{(1)} \r|$, while $\Delta \omega$ are the nearest-neighbor
spacings for the real parts $\omega_k$ of the passive eigenfrequencies. (The
spacings in the figures are scaled with their respective mean values.) $p
\l(\Delta \omega\r)$~for \mbox{$\gamma \ll 1$} is close to the Wigner surmise
\eqn{
  p_W \l(\Delta \omega\r) \approx \frac \pi 2 \, \Delta \omega\, \exp \l(-
  \frac \pi 4 \, \Delta \omega^2 \r),
  \label{pW}
}
which displays eigenvalue repulsion in the GOE. For stronger coupling, the
eigenvalues get spread in the complex plane, and the crossing probability for
the real parts increases. According to the numerical data (Table~\ref{sp}),
the average spacings $\wbar{\Delta \Omega}$ and $\wbar{\Delta \Omega'}$ are of
order of the gain-profile halfwidth~$\gamma_\perp$. We will consider the
three regimes of $\gamma_\perp$ being much greater, of order, or much smaller
than the passive mean spacing $\wbar{\Delta \omega} \approx 1/ L \rho \l(0
\r)$. The inversion decay rate~$\gamma_\parallel$ is taken equal
to~$\gamma_\perp$. 

\begin{table}[tb]
  \begin{tabular}{|c|c||c|c|c|}
    \hline
    $\gamma$ & $\gamma_\perp$ & $\wbar{\Delta \Omega}$ & 
      $\wbar{\Delta \Omega'}$ & $\wbar{\Delta \omega}$ \\ \hline \hline
    & $10^{-3}$ & $0.0049$ & $0.0043$ & \\ \cline{2-4}
    $0.1$ & $10^{-2}$ & $0.017$ & $0.015$ & $0.016$ \\ \cline{2-4}
    & $10^{-1}$ & $0.102$ & $0.079$ & \\ \hline
    & $10^{-3}$ & $0.0020$ & $0.0016$ & \\ \cline{2-4}
    $1.0$ & $10^{-2}$ & $0.017$ & $0.014$ & $0.015$ \\ \cline{2-4}
    & $10^{-1}$ & $0.11$ & $0.087$ & \\ \hline
  \end{tabular}
  \caption{\small Mean frequency spacings~$\wbar{\Delta \Omega}$ between the
  lasing modes ($\wbar{\Delta \Omega'}$ neglecting the mode competition) and
  $\wbar{\Delta \omega}$ between the passive-cavity modes for different
  $\gamma$ and $\gamma_\perp = \gamma_\parallel$.}
  \label{sp}
\end{table}
 
1. $\gamma_\perp \gg \wbar{\Delta \omega}$ [Figs.~\ref{p1}(a)
and~\ref{1p}(a)]. Although the distributions $P \l(\Delta \Omega\r)$ and $P'
\l(\Delta \Omega'\r)$ are quite similar, the mode competition still influences
the lasing frequencies. In principle, this influence is twofold: (1)~the
indices $k = 2$ and $k = 2'$ refer, in general, to two different passive modes
and, (2)~even if the modes are the same, the frequency $\Omega_2^{(2)}$ is
different from~$\Omega_2^{(1)}$ [Eq.~\eq{frshift}]. It can be deduced from
Eq.~\eq{frshift} and was also checked numerically (not shown) that, in the
present case, the second factor is insignificant. On the scale $\Delta \Omega
\gtrsim \wbar{\Delta \Omega}$, the passive distribution $p \l(\Delta \omega
\r)$ is irrelevant for~$P \l(\Delta \Omega\r)$, because the passive modes
corresponding to $k = 1,2$ are not the nearest neighbors. The passive modes
with close~$\omega_k$'s are less likely to become lasing modes, since they may
lie far from each other in the complex plane, i.e., one of them can be
substantially damped. This explains the falloff of~$P \l(\Delta \Omega\r)$ as 
$\Delta \Omega \to 0$. For small~$\gamma$, the repulsion of lasing frequencies
is guaranteed by the Wigner repulsion of the passive modes.

An approximate expression at $\Delta \Omega' \gg \wbar{\Delta \omega}$ for the
distribution $P' \l(\Delta \Omega'\r)$ without mode competition can be derived
from a simple probabilistic model as follows. Of all passive eigenvalues
$\omega_k - i \kappa_k$, the two lasing modes have the smallest $\wtilde
\kappa_k \equiv \kappa_k \l( 1 + \omega_k^2 / \gamma_\perp^2 \r)$ [here we can
approximate $\Omega_k^{(1)} \approx \omega_k$, since only small $\kappa_k$ are
relevant]. Let us divide the $\omega$ axis into $N_\omega \gg 1$ intervals,
each containing $L / N_\omega \gg 1$ eigenvalues. These conditions ensure that
the two lasing modes are unlikely to belong to the same interval and that the
$\kappa_k$ distributions in different intervals are uncorrelated. The smallest
$\kappa_k$ within an interval is distributed according to $P_{\text{gap}} \l(
\kappa \r) = \l(d n / d \kappa \r) e^{-n \l( \kappa \r)}$ [assuming that the
number of eigenvalues between $\kappa$ and $\kappa + d \kappa$ obeys the
Poisson distribution with the average $dn \l( \kappa \r)$]. Here $n \l( \kappa
\r)$, the mean number of eigenvalues with $\kappa_k < \kappa$, has a
small-$\kappa$ asymptotics~$\kappa^{\frac M 2}$~\cite{fyod97} (valid for
arbitrary~$\gamma$). The distribution of the smallest $\wtilde \kappa_k$ is
$\wtilde P_{\text{gap}}^{\omega_i} \l(\wtilde \kappa \r) = P_{\text{gap}} \l(
\kappa \r) d \kappa/ d \wtilde \kappa$, where $\omega_i$ is the central
frequency of the interval. The two lasing modes lie in the intervals $i$
and~$j$ with the probability
\aln{
  P_{12}^{\omega_i, \omega_j} = &\int d \wtilde \kappa\, d \wtilde
  \kappa'\, \wtilde P_{\text{gap}}^{\omega_i} \l(\wtilde \kappa\r) \wtilde
  P_{\text{gap}}^{\omega_j} \l(\wtilde \kappa' \r) \notag \\
  &\times \prod_{l \ne i,j} \l[1 - \int_0^{\max \l(\wtilde \kappa, \wtilde
  \kappa'\r)} d \wtilde \kappa''\, \wtilde P_{\text{gap}}^{\omega_l}
  \l(\wtilde \kappa'' \r) \r].
}
Using an argument similar to that of Ref.~\cite{haak01} (Sec.~5.6), the
product in the second line can be estimated as $\exp \l\{- \mc O \l[ L \max
\l(\wtilde \kappa, \wtilde \kappa'\r)^{M/2} \r] \r\}$. Thus, only $\wtilde
\kappa \ll L^{-2/M}$ is relevant. The spacing distribution~is
\eqn{
  P' \l(\Delta \Omega'\r) \propto \int d \omega\, P_{12}^{\omega, \omega +
  \Delta \Omega'} \propto \int d \omega\,  \l(\mc L_\omega\, \mc L_{\omega +
  \Delta \Omega'} \r)^{M/2},
}
where $\mc L_\omega \equiv \l( 1 + \omega^2 / \gamma_\perp^2 \r)^{-1}$ and the
integration was extended to infinity. For $M=6$ (see the figures),
\eqn{
  P' \l(\Delta \Omega'\r) \approx \frac {32} {3 \pi \gamma_\perp} \frac {x^4 +
  24 x^2 + 336} {\l( x^2 + 4 \r)^5}, \quad  x \equiv \frac {\Delta \Omega'}
  {\gamma_\perp}.
  \label{Pappr}
} 
 
2. $\gamma_\perp \lesssim \wbar{\Delta \omega}$ [Figs.~\ref{p1}(b,c)
and~\ref{1p}(b,c)]. The lasing modes are likely to emerge from the passive
nearest neighbors. Thus, the spacing distribution approaches the Wigner
surmise~\eq{pW}. At small~$\gamma_\perp$, the mode competition is manifested
mostly via the frequency shift~\eq{frshift}, while $k = 2$ and $k = 2'$ is
usually the same mode. This frequency shift is responsible for the vanishing $P
\l(\Delta \Omega\r)$ as $\Delta \Omega \to 0$, even when $P' \l(\Delta \Omega'
\r)$ does not vanish [Figs.~\ref{p1}(c) and~\ref{1p}(c)]. In other words, when
$\Omega_2^{(1)}$ gets close to~$\Omega_1^{(1)}$, $\Omega_2^{(2)}$~is repelled
from these two frequencies. It is expected that the distribution $P' \l(\Delta
\Omega' \r)$ without the mode competition is close to~$p \l(\Delta \omega\r)$.
In particular, when $\gamma_\perp \ll \wbar{\kappa_k} \sim \wbar{\Delta
\omega}$ [Fig.~\ref{1p}(c)], the frequencies~\eq{O1} can be approximated as
$\Omega_k^{(1)} \approx \omega_k \gamma_\perp / \kappa_k$. This makes $\mc
L_k^{(1)} \approx \l[1 + \l(\omega_k/\kappa_k \r)^2 \r]^{-1}$, i.e., the gain
profile becomes independent of~$\gamma_\perp$. When, on the other hand,
$\gamma_\perp \sim \wbar{\kappa_k} \ll \wbar{\Delta \omega}$
[Fig.~\ref{p1}(c)], the two modes with the smallest~$\l|\Omega_k \r|$, not the
smallest~$\kappa_k$, are lasing. Since $\kappa_k$'s for these modes can be
substantially different, $\Delta \Omega'$~is not proportional to~$\Delta
\omega$ [Eq.~\eq{O1}]. This explains the difference between $P' \l(\Delta
\Omega' \r)$ and $p \l(\Delta \omega\r)$ in Fig.~\ref{p1}(c).

\section{Conclusion}

The spacing distribution for the first two lasing modes depends
on the relation between the gain-profile width and the mean spacing of the
passive-cavity modes.  For a very wide gain profile, the passive-spacing
statistics is irrelevant, except for very small spacings.  When these
parameters are of the same order, the spacing distribution is approximated by
the Wigner surmise. The nonlinear frequency shift due to the mode competition
provides for the mode repulsion, even when the unperturbed frequencies may
approach each other.

\begin{acknowledgments}

I thank Carlo Beenakker for his suggestion to study the spacing statistics and
helpful comments. I benefited from discussions with Lev Deych, Fritz Haake, and
Hans-J\"urgen Sommers. Financial support was provided by the Deutsche
Forschungs\-gemein\-schaft via the SFB-TR12.

\end{acknowledgments}

\bibliography{2modes}

\end{document}